*Atmospheric Pressure Mass Spectrometry of Single Viruses and Nanoparticles by Nanoelectromechanical Systems*

R. Tufan Erdogan [1,2,‡], Mohammed Alkhaled [1,2,‡], Batuhan E. Kaynak [1,2,‡], Hashim Alhmoud [1,2,‡], Hadi Sedaghat Pisheh [1,2], Mehmet Kelleci [1], Ilbey Karakurt [1], Cenk Yanik [3], Z. Betul Sen [1], Burak Sari [4], Ahmet Murat Yagci [5], Aykut Özkul [6,7], M. Selim Hanay [1,2,*]

[1] Department of Mechanical Engineering, Bilkent University, 06800 Ankara Turkey

[2] UNAM — Institute of Materials Science and Nanotechnology, Bilkent University, 06800 Ankara Turkey

[3] Sabancı University, SUNUM Nanotechnology Research and Application Center, 34956 Istanbul Turkey

[4] Faculty of Engineering and Natural Sciences, Sabancı University, 34956 Istanbul Turkey

[5] METU MEMS Center, 06530 Ankara Turkey

[6] Faculty of Veterinary Medicine, Department of Virology, Ankara University 06110 Ankara Turkey

[7] Biotechnology Institute, Ankara University 06135 Ankara Turkey

ABSTRACT

Mass spectrometry of intact nanoparticles and viruses can serve as a potent characterization tool for material science and biophysics. Inaccessible by widespread commercial techniques, the mass of single nanoparticles and viruses (>10MDa) can be readily measured by NEMS (Nanoelectromechanical Systems) based Mass Spectrometry, where charged and isolated analyte particles are generated by Electrospray Ionization (ESI) in air and transported onto the NEMS resonator for capture and detection. However, the applicability of NEMS as a practical solution is hindered by their miniscule surface area, which results in poor limit-of-detection and low capture efficiency values. Another hindrance is the necessity to house the NEMS inside complex vacuum systems, which is required in part to focus analytes towards the miniscule detection surface of the NEMS. Here, we overcome both limitations by integrating an ion lens onto the NEMS chip. The ion lens is composed of a polymer layer, which charges up by receiving part of the ions incoming from the ESI tip and consequently starts to focus the analytes towards an open window aligned with the active area of the NEMS electrostatically. With this integrated system, we have detected the mass of gold and polystyrene nanoparticles under ambient conditions and with two orders-of-magnitude improvement in capture efficiency compared to the state-of-the-art. We then applied this technology to obtain the mass spectrum of SARS-CoV-2 and BoHV-1 virions. With the increase in analytical throughput, the simplicity of the overall setup and the operation capability under ambient conditions, the technique demonstrates that NEMS Mass Spectrometry can be deployed for mass detection of engineered nanoparticles and biological samples efficiently.

### Introduction

Nanoelectromechanical Systems (NEMS) are powerful tools capable of performing mass spectrometry (MS) on large analytes with masses typically > 10 MDa.[1-9] As such, they cover a mass range that is not accessible by conventional MS systems due to large mass-to-charge ratios (m/z) of analytes.[10-12] The analytes in this mass range include metallic, ceramic, and polymeric nanoparticles, in addition to quantum dots, large supramolecular assemblies, exosomes, viruses, and lipid vesicles, all of which have been so far outside the realm of commercial MS instruments. Only one MS technique, Charge Detection MS[13-16] (CDMS) has been demonstrated to work at large mass values, spanning between several kDa to hundreds of MDa in mass range. CDMS works by carrying out a simultaneous measurement of the analyte charge in addition to its mass-to-charge (m/z) in order to derive the mass of the analytes,[17, 18] recently attaining high resolving power (>300) for species at several MDa range.[19, 20] Unlike CDMS and any other MS techniques that utilize m/z detectors, NEMS technology can directly measure the mass without any dependence on the charge state of the ions. This renders NEMS-MS immune to any discrepancies in the spectrum caused by identical species with different charge states which is regarded as an especially important advantage for characterizing nanoparticles.[21] Another salient feature of NEMS-based sensors is that they have a small form factor (since NEMS is a chip-based technology), potentially allowing them to be embedded in compact instruments leading to true portability and miniaturization.

Paradoxically, this small form factor is one of the major drawbacks preventing NEMS-based Mass Spectrometry (NEMS-MS) from ubiquitous adoption as a fully realized MS technique. To perform NEMS-MS, analyte particles need to be transported to a NEMS beam vibrating at the resonance frequency. The resonating beam is typically a few micrometers long and several hundred nanometers wide. The vast majority of analyte particles accelerated towards the NEMS device land predominantly outside of the NEMS capture area (*i.e.* resonating beam area) and therefore do not register as landing events. This translates into a low analyte capture efficiency — the ratio of analyte particles detected by NEMS over the total particles used during the analysis. To ameliorate this limitation, researchers resorted to using ion optics[1, 3] or aerodynamic lensing[5, 6, 22] to transport and focus the stream of analytes emitted from a soft ionization system, such as Electrospray Ionization (ESI), towards the NEMS capture area in order to detect a sufficient number of analyte particles within a reasonable time frame. However, both ion optics and aerodynamic lensing in NEMS-MS systems require differential vacuum systems to operate. Differential vacuum systems are composed of multiple vacuum components leading to a delicate design, thus increasing the cost, size, and complexity of the system. Therefore, the whole instrument becomes encumbered by the vacuum setup, leading to a loss in its apparent advantages (miniaturization, portability, feasibility of mass-production *etc.*). In fact, even with the most optimized designs, the current state-of-the-art capture efficiency[5, 6] is still worse than 1 particle out of 20 million contained in the original solution which is not sufficient for analyzing real clinical samples.[23, 24]

In order to eliminate the vacuum requirement while providing superior focusing performance, we have designed an architecture for NEMS-MS that can operate under atmospheric conditions, where the focusing of the analytes is accomplished on-chip (Figure 1a). On-chip ion focusing plays a critical role, since standard ion optics techniques do not work at atmospheric conditions:[25, 26] rather we were inspired by recent advances in nanopatterning[27] and microscale 3D printing[28] to form an integrated system. The on-chip ion lens is composed of a photoresist layer with an open window exposing the NEMS resonator. The photoresist layer charges up by accumulating part of the incoming ions on itself and establishes ion focusing onto the NEMS capture area (Figure 1b). This way, it works as a self-biasing lens with the focusing region perfectly aligned with the NEMS capture area.

With this system, we have demonstrated efficient characterization of different nanoparticles and viruses in relatively low concentrations (< $1\times10^{10}$ Particles/mL) under atmospheric pressure with analysis time of typically 30 minutes. To our knowledge, atmospheric MS in this context was never achieved before. The integrated lens increased the capture efficiency of gold nanoparticles (GNPs) by several orders of magnitude exceeding the capture efficiencies reported by state-of-the-art NEMS-MS architectures in the literature. Furthermore, polystyrene nanoparticles were successfully focused and detected by the NEMS-MS device also under atmospheric pressure and with record capture efficiencies. Finally, as a proof of utility in the biomedical field, the NEMS-MS system as described was able to measure the mass of SARS-CoV-2 virions from cell lysate in addition to bovine herpes virus (BoHV-1). We believe that the development of this integrated lens approach will solve a major hurdle preventing the nanoscale mass detectors from reaching their true potential of high capture efficiency, atmospheric operation, versatility, and portability.

**Results/Discussion**

**Nanoelectromechanical Sensors with Integrated Lenses**

The NEMS device is a suspended, doubly-clamped silicon nitride beam on a silicon substrate which vibrates mechanically in air. Standard nanofabrication procedures were used to fabricate the beam (10-15 µm long, 400-800 nm wide and 100 nm thick, typically) and deposit gold electrodes on the beam's top surface for integrated electronic transduction.[29, 30] The suspended devices were then spin-coated with a high-viscosity photoresist (AZ 4533) to a final layer thickness of 3.5 µm. After the deposition of the photoresist, the NEMS chip was exposed to ultraviolet (UV) light through an aligned photomask, followed by a development step to open a window in the photoresist layer that is aligned with the NEMS resonator (Figure 1a). This way the photoresist surrounds the NEMS structure and forms a polymeric lens to direct the incoming analytes towards the active sensing area of the NEMS device. The polymeric focusing lens window on our device architecture is typically 20 microns long by 15 microns wide.

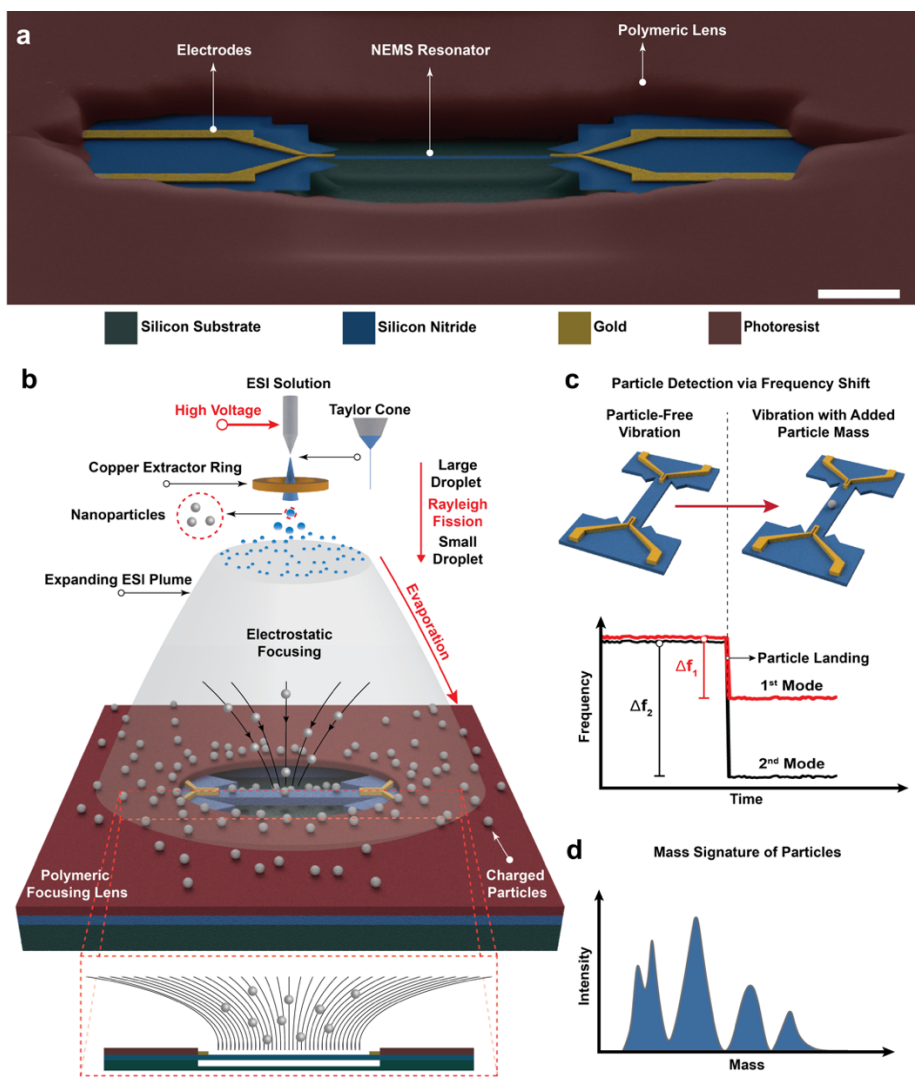

**Figure 1.** Device architecture and the overall scheme of the NEMS-MS system. **(a)** A colorized Scanning Electron Microscope (SEM) image of a NEMS device and the surrounding polymeric lens. The transduction electrodes are shown in yellow. Scale bar is 5 μm. **(b)** Charged droplets containing the nanoparticles or viruses are produced by electrospray ionization (ESI). The charged droplets evaporate and undergo Rayleigh fission on their way to the NEMS chip, yielding single

nanoparticles/virions. The NEMS chip is comprised of the active NEMS device and a photoresist layer having an opening aligned with the active device. The photoresist layer charges up by accumulating the incoming ions, and charges in such a way that it deflects a significant portion of the incoming ions towards the open window, effectively focusing them on the NEMS active area. The illustration is not to scale. **(c)** The incoming particles/virions are then detected by a NEMS device resonating simultaneously in the first two flexural, out-of-plane modes so that the mass and position of each incoming particle are calculated. **(d)** The measured mass spectrum can be used for species identification.

The NEMS chip is placed in a custom setup which involves an electrospray ionization (ESI) subsystem that generates individual nanoparticles, an extractor lens, and a printed circuit board (PCB) to hold the NEMS chip itself (Figure 1b). The entire system is operated at atmospheric pressure in contrast with earlier analytical NEMS instruments housed in vacuum. The analyte solution is introduced into the ESI tip by a syringe pump, and a high electrical potential (typically 5.5 kV) is applied to the solution to induce the formation of a Taylor cone at the tip (Supporting Figure S1). A digital microscope was used to continuously monitor the ESI tip for proper operation. The extractor lens is held at a voltage of 1.3 kV. Charged droplets are emitted from the tip and undergo evaporation and Rayleigh fission to yield single analyte ions along their trajectory to the NEMS chip which is typically located 15 cm away (Supporting Information). The distance between the ESI tip and the NEMS chip, and the ESI parameters are adjusted so that the evaporation of sub-micrometer droplets and desolvation of the analyte particles are ensured, as experimentally verified in Supporting Information.

In order to optimize the ESI parameters such as the operation voltage and the distance between the ESI tip and target chip, blank silicon chips with integrated polymeric lenses were first used in ESI experiments (Figure 2a-d). In these experiments, 100 nm fluorescent polystyrene nanoparticles (Fluoro-Max G100) were used, since the fluorescence of the nanoparticles allowed for the visual observation of the position and density of nanoparticles deposited on the chip using fluorescence microscopy. Using these chips, highly efficient focusing of single nanoparticles was observed. Most of the incoming charged particles and ions accumulated on the photoresist surface initially, however, as more charged particles accumulated, the steady-state build-up of electrostatic charge caused the particles to deflect towards the open window on the polymeric lens. Importantly, under the operation conditions no microdroplets were observed on the surface of the chip. Additionally, almost all of the nanoparticles on top of the polymeric layer, silicon substrate and the NEMS beam are observed as single particles which indicate that the nanoparticles are well separated from each other after ESI while they are traveling towards the chip. In an unoptimized case where the evaporation of microdroplets is not fully completed, the clusters of particles can be observed within the target chip as the microdroplet is deposited on the surface and evaporates slowly, leaving an easily recognizable residue (Supporting Information).

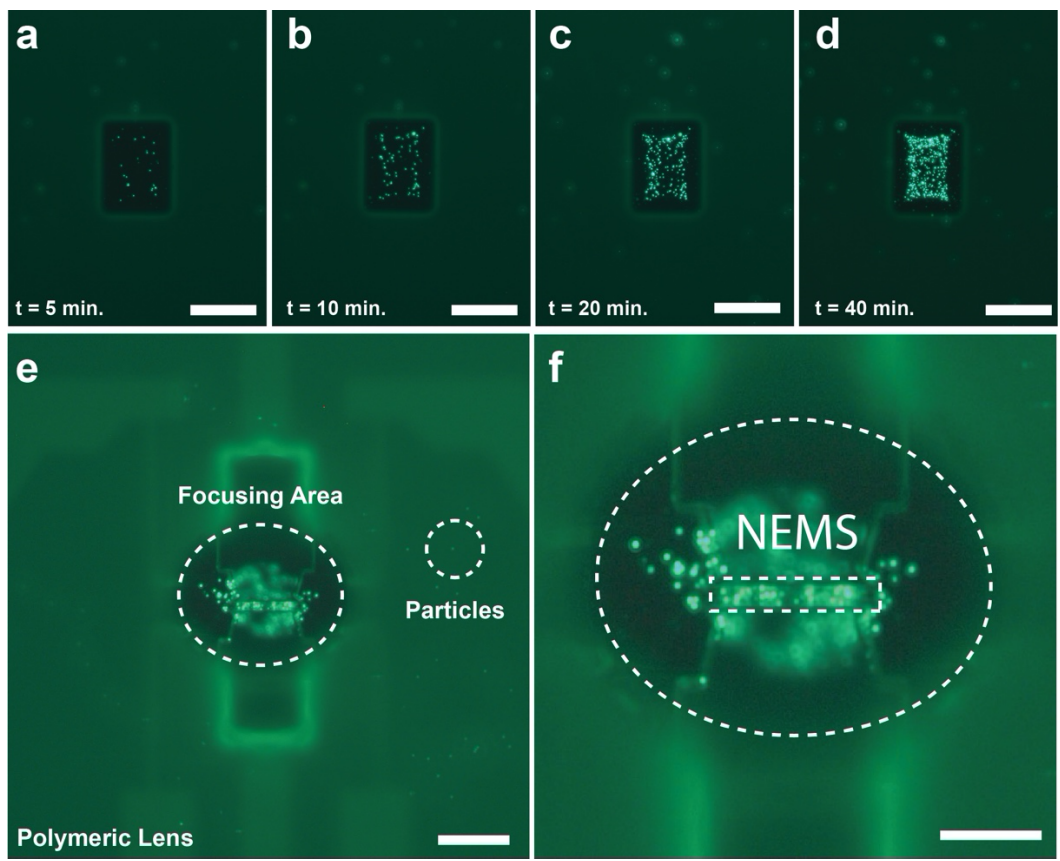

**Figure 2.** Focusing experiments. **(a)**, **(b)**, **(c)** and **(d)** Fluorescence microscopy images of accumulated particles inside a polymeric lens integrated with a bare silicon chip (*i.e.* without the NEMS device) after a total ESI duration of 5, 10, 20 and 40 minutes respectively. Scale bars are 30 μm. **(e)** Fluorescence microscopy image showing the focusing of nanoparticles on the NEMS device after an ESI experiment. The large oval dashed line shows the edge of the polymeric lens; the small dashed circle shows a single nanoparticle on the photoresist. There are dozens of single particles on the entire image, most of them are located within the boundaries of the focusing window. A portion of the particles inside the window are out of focus since they land at the silicon substrate which is at a lower depth than the NEMS surface. Scale bar is 15 μm. **(f)** zoomed in fluorescence microscope image of the NEMS area active area indicated by a dashed line. Scale bar is 10 μm.

After establishing the focusing ability of the polymeric lens on blank silicon chips, and optimizing the process parameters, we then used NEMS devices with integrated lenses to verify that self-focusing still takes place (Figure 2e). Critically, a considerable portion of the particles arriving at the focusing window were further focused and concentrated on the surface of the beam due to the tight alignment of the focusing lens with the NEMS device (Figure 2f).

Next, we conducted experiments to quantify the enhancement of capture efficiency resulting from the integration of the focusing lens with the NEMS devices (Figure 3). Two identical NEMS devices were used, one with and the other without an integrated polymeric lens and compared the number of gold nanoparticles (GNP) detected by each NEMS under the same conditions. 40 nm diameter gold nanoparticles (Sigma-Aldrich 741981) with a concentration of $7.15 \times 10^9$ particles per mL (80 mM ammonium acetate buffer, pH 7.1) were used for this investigation. For the NEMS device without a focusing lens (Figure 3a), a total of 29 particles were registered as landing events by NEMS after 170 minutes of deposition. Also, Scanning Electron Microscopy (SEM) analysis revealed that the particles had a uniform distribution across the entire chip: consequently, the probability of a particle landing on the beam, and an event being subsequently detected by NEMS was low (Figure 3b-c). However, with the focusing lens (Figure 3d), after only 5 minutes of ESI, a total of 101 particles were registered as landing events by NEMS. The particle density on the beam area was much higher compared to the entire chip (Figure 3e-f), as the on-chip polymeric lens successfully focused the incoming particles towards beam surface. In terms of capture efficiency, the NEMS device with the integrated lens registered 1 out 177,000 from the particles overall, while the bare NEMS device (*i.e.* without an integrated lens) registered only 1 out of 21 million particles overall. The presence of the focusing lens increased the rate of transportation of the particles on the NEMS resonator by more than two orders of magnitude.

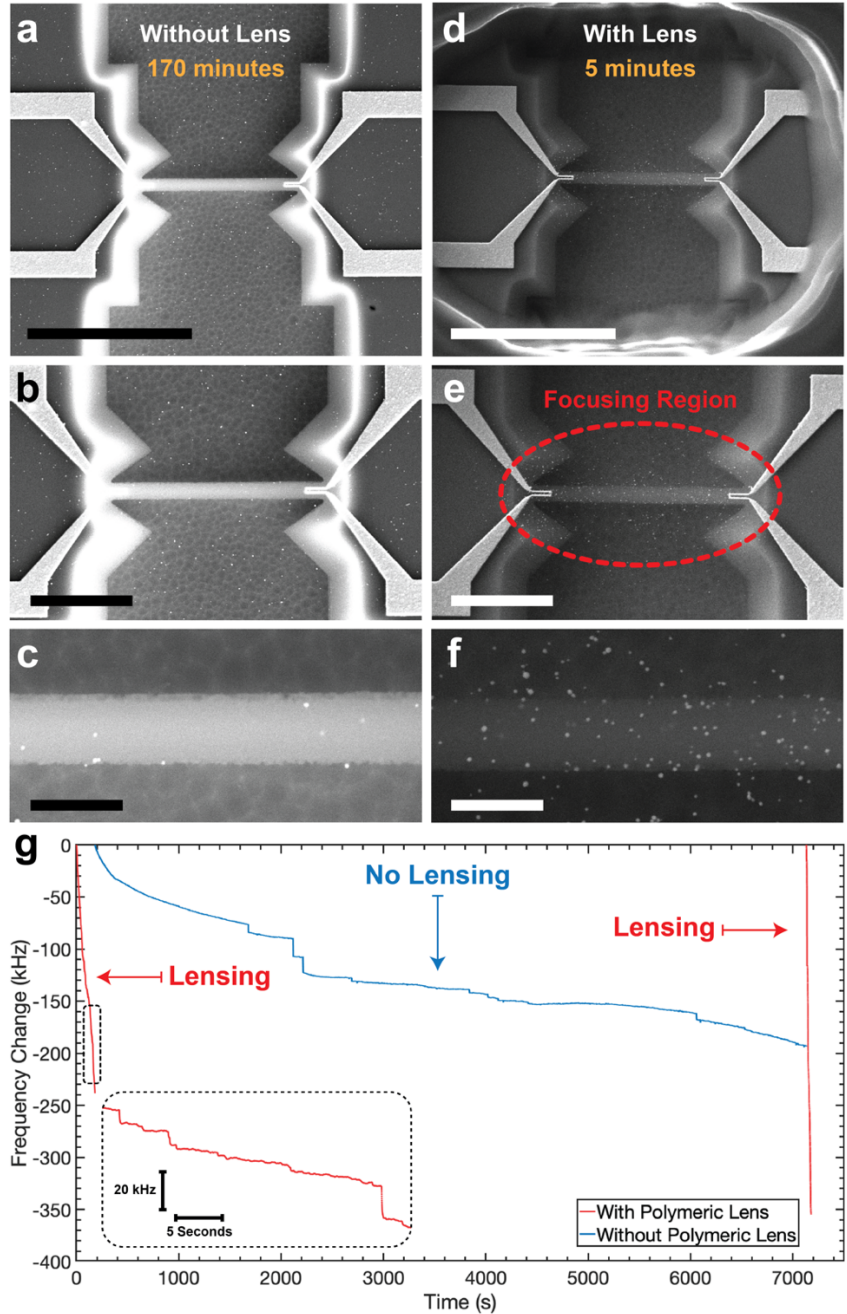

**Figure 3.** The comparison experiments for non-lensing *versus* lensing devices. **(a)** The SEM image of the device without polymeric lens after 170 minutes of ESI. The incoming particles landed everywhere uniformly including the surface of the beam. **(b)** shows a zoomed in SEM image of the device without lens. **(c)** shows an area on the NEMS beam and landed nanoparticles on the device in **b**. **(d)** SEM image of the device with the polymeric lens after 5 minutes of ESI. The focusing successfully takes place with the help of polymeric lens and the incoming particles are concentrated on the beam area. **(e)** shows a zoomed in SEM image of the device in **d**. Deposition region of nanoparticles is indicated with a red dashed line. **(f)** shows an area on the NEMS beam and landed nanoparticles on the device in **e**. The number of particles on the beam is much larger on the device with the integrated polymeric lens compared to the device without lens despite the great difference in ESI duration. **(g)** PLL of the fundamental resonance mode of two different devices that were used in consecutive comparison experiments. The NEMS device with lensing registered more events over a much shorter time interval (note the large change in frequency). Inset shows a zoom-in of one PLL data set for the lensing device where consecutive landing events can be observed. Scale bars are 10 μm for **a** and **b**, 5 μm for **c** and **d**, and 1μm for **e** and **f**.

The on-chip focusing trend observed with the SEM imaging was simultaneously verified by the Phase-Locked Loop (PLL) operation of each device (Figure 3g). The decrease in the experimental duration and the drastic increase in the capture efficiency eliminates the need for aerodynamic lenses and vacuum components. Therefore, the on-chip focusing lens allows for the acquisition of a high number of particles at a given sample concentration in a much shorter time duration compared to the state-of-the-art NEMS-MS where several hours are required to characterize a statistically significant number of incoming analytes.[6]

**Nanoparticle Measurements**

After establishing the focusing ability of the polymeric lens, we tested the performance of our NEMS-MS system using 40 nm and 20 nm gold nanoparticles, and 100 nm polystyrene nanoparticles (80 mM ammonium acetate buffer, pH 7.1, for further details see Methods). We first conducted experiments using 40 nm nominal diameter spherical gold nanoparticles (Sigma-Aldrich 741981). The concentration of the particles in the solution was $7.15\times10^9$ particles per mL (Methods). Figure 4 shows the two mode PLL data of a 40 nm gold nanoparticle detection with a mass resolution of ~3.3 MDa (Supporting Figure S7). This reported mass resolution is suitable for resolving nanoparticles at the targeted range (typically > 50 MDa). The inset in Figure 4 shows two consecutive landing events registered by the NEMS device. While there is a slow drift in the PLL data, it does not significantly obscure the sharp frequency jumps induced by single particles on NEMS (other than slightly increasing the effective noise, *e.g.* Supporting Figure S7b), which is a general characteristic of NEMS resonance detection.[3]

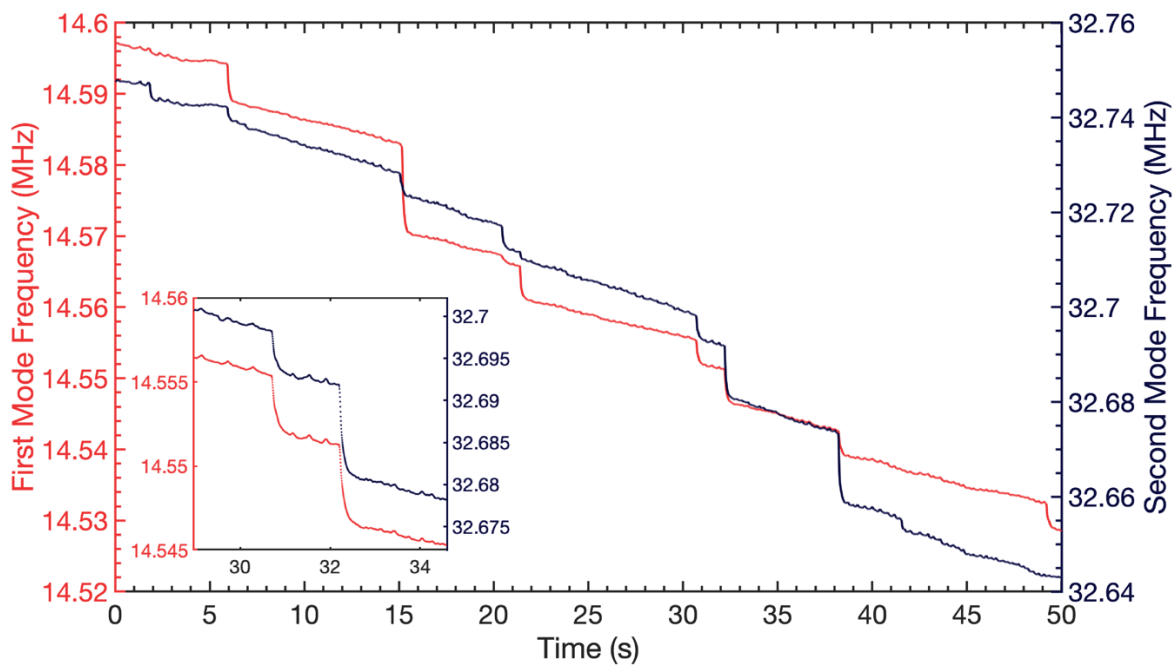

**Figure 4.** Two mode PLL data of 40 nm GNP with mass resolution of ~3.3 MDa. Inset shows a zoom-in of two consecutive landing events.

Figure 5a shows the acquired mass histogram with NEMS-MS for the 40 nm GNP experiments. To verify our NEMS based mass measurements, we set out to compare the results with common sizing techniques (SEM and DLS), by converting the mass histogram into a size histogram using the density of gold (19.27 g/cm$^3$) and assuming a spherical shape, (Figure 5b). The size histogram was in good agreement with the SEM measurements of the GNPs collected at the NEMS beam surface in the same experiment (Figure 5b). Fitting the NEMS size histogram to a non-parametric density function we obtained a global maximum at 35.6 nm (~274 MDa, Supporting Figure S9). Once the size measurements by NEMS and SEM were shown to agree with each other, we then compared these results with dynamic light scattering (DLS). DLS showed a peak value higher than the actual diameter measurements obtained by SEM and NEMS-MS, as this technique probes the hydrodynamic diameter of the gold nanoparticles (Figure 5b) and yields results typically larger than the actual GNP diameter due to the changes in the intensity of scattered light as a function of particle diameter, solution concentration, and moieties adsorbed at the surface of the nanoparticles.[31, 32]

To test the versatility of the atmospheric NEMS-MS, we repeated the same mass measurements on nominally 20 nm GNPs and 100 nm polystyrene nanoparticles (PSNP) at concentrations of $6.54 \times 10^9$ and $3.55 \times 10^9$ particles/mL respectively. Mass measurement peak values of 71 MDa (22.7 nm) and 400 MDa (106 nm, 1.04 g/cm$^3$) were obtained for GNPs and PSNPs respectively. The derived size histograms corresponded closely to validation measurements by both SEM and DLS as is shown in Figure 5c-f. In the measurements of both gold and polystyrene nanoparticles, we have obtained capture efficiencies several orders of magnitude better than the state of the art (Supporting Information) owing to on-chip ion focusing.

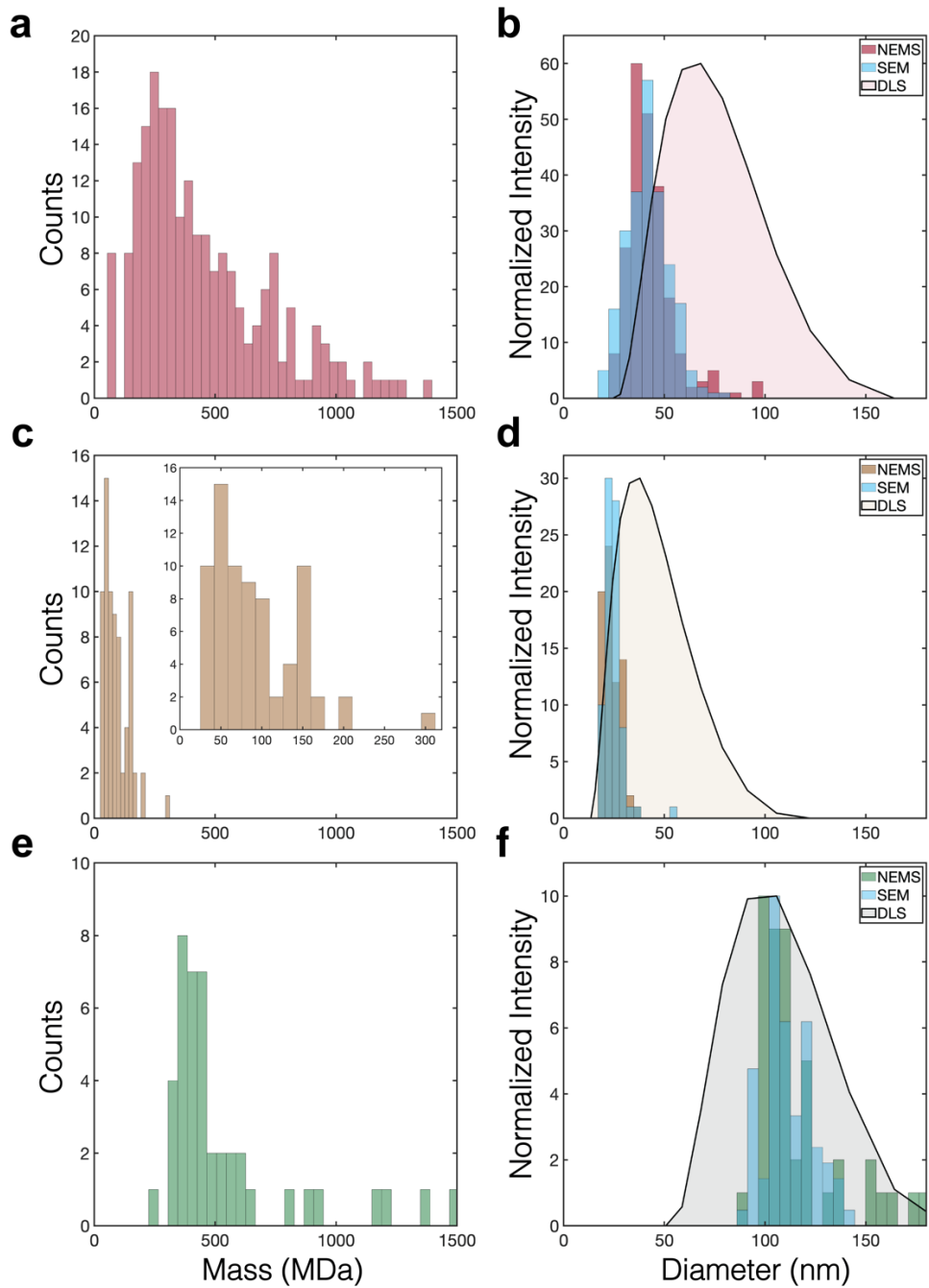

**Figure 5.** Nanoparticle measurements. **(a)** 40 nm GNP mass histogram acquired with NEMS-MS **(b)** NEMS-MS, SEM and DLS diameter comparison for 40 nm GNP. **(c)** 20 nm GNP mass histogram acquired with NEMS-MS. Inset shows the same histogram for clarity **(d)** NEMS-MS, SEM and DLS diameter comparison for 20 nm GNP. **(e)** 100 nm PSNP mass histogram acquired with NEMS-MS **(f)** NEMS-MS, SEM and DLS diameter comparison for 100 nm PSNP. For each subfigure, the y-axis is counts for the NEMS-MS and SEM measurements, and normalized intensity for DLS measurements.

**SARS-CoV-2 Measurements**

Once the system was benchmarked using gold and polystyrene nanoparticles, we switched to biological samples. A SARS-CoV-2 sample, isolated from a COVID-19 patient, and expanded *in vitro*, was analyzed first. The virus sample was harvested by applying a freeze-thaw cycle twice to lyse the cell culture. The cell lysate was then centrifuged to remove cell debris, followed by a short thermal inactivation for safety purposes (Methods, SARS-CoV-2 Virus Isolation). Before the thermal inactivation, SARS-CoV-2 samples had a concentration of approximately $2\times10^6$ PFU/mL as calculated by a viral plaque assay. The virus was isolated using PEG precipitation followed by dialysis at 4°C to exchange the culture media with the ESI-compatible 80 mM ammonium acetate solution (pH 7.1). Dynamic Light Scattering (DLS) was then conducted on the sample to obtain the hydrodynamic diameter of the isolated virus in the sample and to confirm the absence of any other major particulates left over from the cell lysate. DLS characterization showed one major peak centered at 86 nm, assumed to mostly correspond to the virion particles, while no other major peaks were detected. The same sample, with virions diluted forty-fold, was then analyzed using NEMS-MS, resolving 153 single virion events within 14 minutes. The resulting mass spectrum is shown in (Figure 6a) which was used to fit a non-parametric probability density function. The global maximum value for the density function is 454 MDa.

To interpret this spectrum, first, we converted the mass values into effective diameter values, assuming that the SARS-COV-2 virions had a spherical shape and using a typical value for the viral mass density (1.4 g/cm$^3$).[33] After the conversion, the resulting diameter histogram was generated as shown in (Figure 6a inset), together with the DLS result of the same sample. The peak value for the mass distribution (454 MDa) corresponded to a diameter of 99.7 nm, while the mean diameter was 101.8 nm with a standard deviation of 13.1 nm. The size of SARS-CoV-2 particles has been recently reported[34-36] by different microscopy techniques (Supporting Table S6) which show that these virion particles have a mean diameter in the range of 90-97 nm with standard deviations ~10 nm. Most of the aforementioned studies report the size values without the peplomers (*i.e.* the spike proteins projecting outwards from the viral envelope), while NEMS measurements include the entire structure: hence, it is expected that the size reported by NEMS-MS is slightly larger. The similarity of the values indicates that the NEMS-MS system, operating at room temperature and atmospheric pressure, can characterize viruses as in the case for nanoparticles based on their mass.

After this additional verification, PEG precipitation was removed from the protocol to check the performance of the NEMS on a relatively complex biological solution such as the clarified cell lysate. The cell lysate was subjected to dialysis in order to remove any non-volatile salts for ESI. The DLS analysis of the lysate showed a peak near 115 nm coalescing with a much larger peak spanning the 200-1200 nm range presumably owing to the interference of cellular debris in the sample. Spraying such a solution containing relatively massive particulates can potentially damage the suspended NEMS beam. This was what we observed in several runs, where the device stopped responding after a few minutes of ESI potentially due to damage by the landing of relatively massive particles. This can be remedied by centrifuging the cell lysate for longer or at a higher speed than in the described protocol to remove the majority of the larger cell debris. On such a trial, a total of 101 events were registered after approximately 18 minutes of spraying without damage to the beam. Most of these events fell on the mass range determined from the isolated SARS-CoV-2 sample (Figure 6b). Compared to the isolated virus sample, there were additional particulates detected especially at the higher mass values, presumably originating from the cellular debris. However, above this background, a clear, population emerged at the expected location for SARS-CoV-2 with the major peak almost overlapping with the peak of the spectrum of the isolated virus (469 *vs.* 454 MDa respectively, for further statistics, see Supporting Table S5).

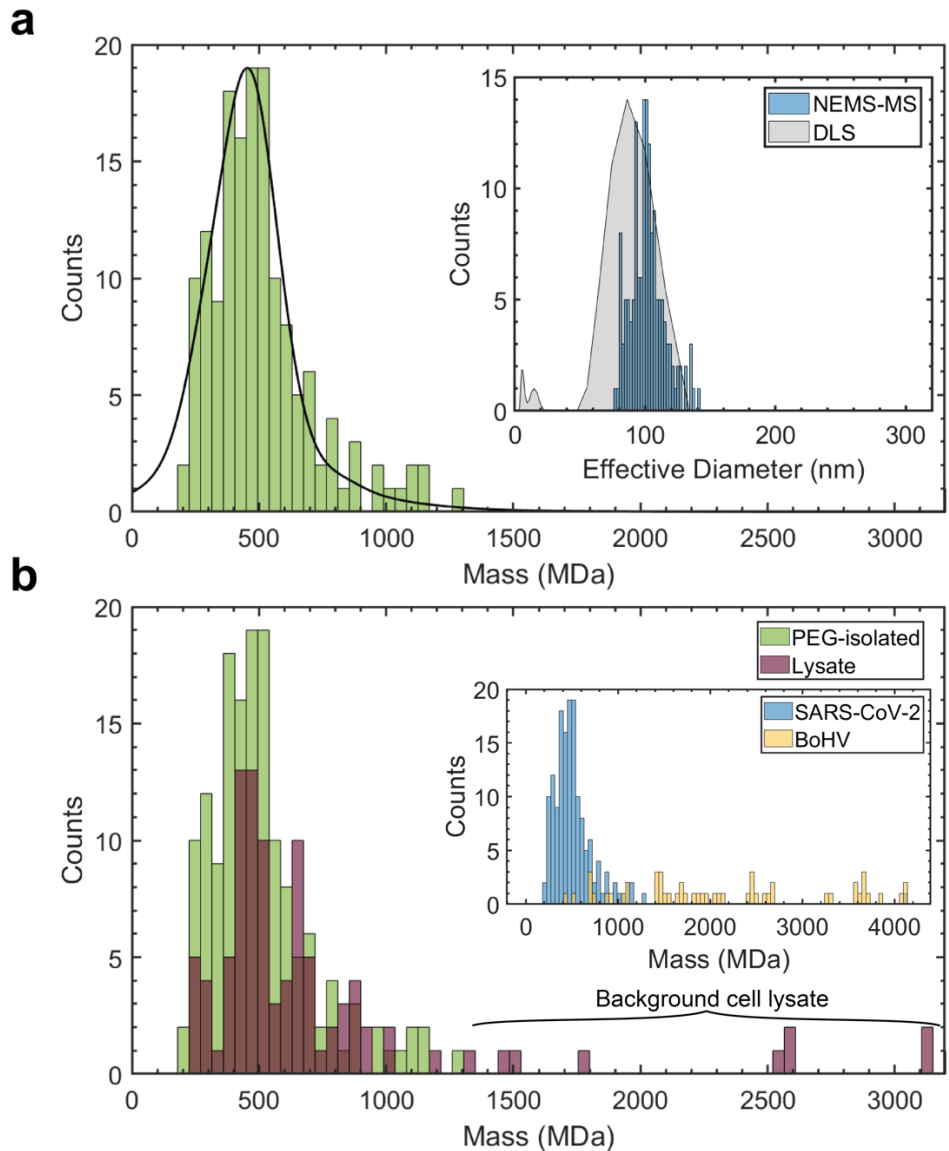

**Figure 6.** Single SARS-CoV-2 virion mass sensing. **(a)** The mass spectrum accumulated after 153 landing events for SARS-CoV-2. A non-parametric density function is fitted to the histogram. Inset shows the corresponding effective diameter histogram calculated by the assumption that the virus density is 1.4 g/cm$^3$ shown together with the DLS analysis result for the same sample. **(b)**

Green: the mass spectra histogram of a SARS-CoV-2 lysate obtained after isolating the virus using PEG precipitation. Red: the mass spectra density function of the non-isolated sample. Inset shows comparison of positive control experiment of BoHV-1 sample. For the inset: there are 153 events for SARS-CoV-2 and 52 events for BoHV-1. For BoHV-1, 31 more events were registered with mass larger than 4.2 MDa.

Furthermore, we analyzed an inactivated Bovine herpesvirus (BoHV-1) sample, processed in the same manner as the isolated SARS-CoV-2 sample using PEG, to both act as a control experiment and show the capability of the system in detecting virions from a different mammalian virus family. This virus has a more complex structure compared to SARS-CoV-2 and has a larger mean diameter. Moreover, the virion may incorporate more than one nucleocapside which results in a broader size distribution as shown by electron microscopy.[37] The NEMS-MS results in this case agreed with the expected size distribution of BoHV-1 (Figure 6b inset). More importantly, specific peaks near the expected mass and size values that emerged from the SARS-CoV-2 analyses were not present in the run conducted with BoHV-1.

It was not possible to measure the actual virion concentration prior to ESI, since the standard technique of plaque assay quantifies only the active viruses. Therefore, assessing the capture efficiency achieved through NEMS detection for the particular case of SARS-CoV-2 was not possible. Nevertheless, this work clearly establishes that NEMS-based Mass Spectrometry can now be used in more realistic and relevant biological samples with concentrations as low as $5\times10^4$ PFU/mL (and possibly lower due to losses during sample processing), and sample volumes as little as 7 μL. Moreover, this analysis was also done directly on complex media such as a cell lysate which eliminates the necessity for complex sample preparations. Therefore, the mammalian viruses such as SARS-CoV-2 can be detected and characterized based on its mass using NEMS-MS entirely under atmospheric conditions. Also, accomplishing the characterization of the viruses is very similar to the characterization of other nanoparticles owing to the high masses detectable under ambient conditions and over short durations.

**Outlook and Perspective**

In this work, we have established a high-throughput system operating at ambient conditions for the direct mass measurements of individual particles and viruses. Although we used electrospray ionization to facilitate the transportation and focusing of the analytes, the NEMS MS technique does not need the charge state of the analyte to deduce its mass since the resonance frequency of the NEMS beam is a function of the mass but not the charge. Unlike recent trends for the use of neutral species,[4, 6] in this work we also show that the charged analyte particles, formed by the well-established ESI process, can be used as a resource to electrostatically bias the photoresist lens leading to ion focusing and an increase in capture efficiency.

Before the use of ESI,[1] earlier work with MEMS/NEMS technology at ambient conditions such as the detection of E. Coli cells[38] and viruses,[39] were conducted by incubating the devices with analytes in liquid, and then drying the chips and measuring the resonance curves in air. However, due to the long duration involved for analyte capture (*via* diffusion) and drying process of the chips, as well as higher uncertainties involved in the measurements, these approaches did not evolve into automated, high-throughput systems. By contrast, the approach taken here, where ESI is utilized, resolves the capture and wetting problems by allowing the sprayed droplets to completely evaporate on their way to the chip-based resonator beam. Another approach to operate NEMS devices is the use of aerosol impaction.[22] In this approach, analyte nanoparticles are generated in the gas phase by first electrospray ionization, followed by neutralization of the nanoparticle charge in an intermediate chamber. The neutralized particles in this chamber are then aerodynamically transported through a narrow aperture where a NEMS device is placed. Since the neutralization of the nanoparticles, and their transportation in between the different chambers of the system results in inevitable sample losses, we have instead concentrated on directly working with the charged particles and focus them using an ion lens.

Since NEMS-MS measures the mass of landing analytes directly (without using charge as a proxy), it avoids some of the pitfalls encountered by conventional MS where the overcrowding of different mass-to-charge peaks prohibits the usage of conventional MS for species heavier than ~10 MDa. By measuring the charge and the mass-to-charge ratio of individual ions separately, species larger than 10 MDa can be measured by the Charge Detection Mass Spectrometry approach.[13-16] While some studies with this instrument accomplished the detection of structures larger than 100 MDa,[17, 18] the practitioners of the field often regard NEMS technology as a more suitable alternative for operation at this extremely large mass range.[21] Additionally, as with other MS techniques, Charge Detection MS requires vacuum instrumentation, in the form of differential vacuum system, which increases the cost, size and the maintenance requirements. On the other hand, the NEMS MS with focusing polymeric lens does not require any vacuum components.

Apart from mass spectrometry approach, there are other approaches to determine the size (but not the mass) of large biological structures and nanoparticles such as IMS (Ion Mobility Spectrometry) and DLS. IMS measures the mobilities of ionic species and then uses an empirical correlation curve to relate the ion mobility to particle mass.[40, 41] However, the correlation curve is not ideal: for instance, the shape and type of the analyte as well as its charge state affects the mobility significantly.[42, 43] As such the mass values extracted from IMS are taken as indicative rather than actual mass values. Moreover, IMS systems require delicate instrumentation involving air flows, increasing the size and cost of IMS equipment in a similar fashion to earlier NEMS-MS designs[3, 5, 6] but unlike our chip-based atmospheric pressure NEMS-MS technology.

Another commonly used tool for size (hydrodynamic radius) characterization of nanoparticles is DLS which suffers from limitations especially in complex mixtures.[44] The scattering cross-section depends on the shape and material properties of the particles; and the signal intensity depends on the diameter of the particle with a sixth order dependency, creating misrepresentations in polydisperse samples. Moreover, parameters such as the solvent viscosity, surface functionalization *etc.* need to be known for correct calibration of the instrument. We also note that the results show that the NEMS-MS can be both more accurate and more precise than DLS. A more detailed discussion of NEMS-MS technology with other related techniques such as Mass Photometry,[45] thermophoresis,[46] and Gas-phase electrophoretic molecular mobility analyzers (GEMMA)[40] is provided in the Supporting Information.

After evaluating the competing technologies, our self-focusing on-chip NEMS-MS system has been shown to provide clear and tangible advantages over current mass and size detection technologies that have been discussed. The major advance brought forward by our design is the elimination of vacuum requirements resulting in the miniaturization of the mass detector. Moreover, the integration of a photoresist focusing layer increases the capture efficiency of NEMS sensors by two orders-of-magnitude which decreases the analysis time significantly. In our opinion, these advances are critical components for the development of first generation commercial NEMS-MS systems that can be either portable or benchtop and that are capable of providing rapid analysis either at point-of-care or in research and development in the biomedical field. For such applications, the limiting factor is to miniaturize and accelerate the pre-processing protocol of samples, either by optimizing established methods (*e.g.* ultracentrifugation for buffer exchange) or by utilizing emerging techniques, such as deterministic lateral displacement[47] or asymmetrical flow-field fractionation.[48] The capabilities of such a system will also extend its utility to other disciplines such as the environmental sciences for pollution detection and air quality control.

**Conclusions**

The integration of the ion lens allowed us to operate our entire system under ambient conditions with a mass resolution of ~ 3.3 MDa. The obtained mass resolution may seem modest, but for the large mass range (> several 100 MDa) considered for nanoparticle and virus applications, it is small enough to distinguish between two particles within this mass range. More importantly, the mass spectrum of engineered nanoparticles and viruses exhibit an inherent dispersion (*e.g.* because of the synthesis conditions, and the stochastic nature of virion formation including the number of spikes in the case of SARS-CoV-2). Therefore, the mass resolution attained under ambient conditions appears to be suitable to differentiate different species (*e.g.* by Figure 5 and Figure 6b inset) owing to the large mass values for the species targeted, as well as the inherently polydisperse mass distribution for each analyte.

In this work, we have demonstrated mass spectrometry measurement of single nanoparticles at atmospheric conditions and obtained capture efficiencies several orders-of-magnitude larger than the earlier state-of-the-art, and with adequate mass resolutions by the on-chip integration of NEMS sensors with self-biased polymeric lenses. The integration of on-chip lensing with NEMS enhanced the capture efficiency and reduced the analysis time by several orders-of-magnitude. With this system, we have characterized different classes of nanoparticles and viruses, with an analysis time of less than thirty minutes. As this form of mass spectrometry operates entirely at ambient conditions, the constraints imposed by vacuum instrumentation are circumvented, which contributes toward the development of miniature and field-deployable systems based on nanoscale sensors.

**Methods/Experimental**

**Details of Micro/Nanofabrication.** The NEMS sensor fabrication process consists of two main steps: first, the fabrication of a suspended doubly-clamped beam integrated with the transduction electrodes, and second, the creation of the electrostatic polymeric focusing lens aligned with the NEMS device.

The NEMS sensor was fabricated from a 100 nm thick stoichiometric Silicon Nitride film on a 500 μm Silicon substrate (University Wafer). First, the metallization of the actuation and read-out electrodes along with the contact pads was performed. To this end, electron beam lithography (EBL) was used with Poly (methyl methacrylate) (PMMA) to pattern the electrodes. After PMMA development, 5 nm chrome adhesive layer and 70 nm gold layer was deposited using thermal evaporation. Then a second EBL process was carried out to define the mechanical structure consisting of the beam and clamping region. This was followed by a 40 nm copper deposition using thermal evaporation to be used as a hard mask in the following steps to suspend the beam. To suspend the beam, an anisotropic dry etch for Silicon Nitride, followed by an isotropic dry etch for Silicon was performed. Finally, the copper hard mask was removed using an isotropic wet etch. The NEMS sensor fabrication is described in detail in references.[30, 49]

To create the polymeric lensing structure, a second major process was developed. We used a commercial photoresist to constitute the polymeric layer, due to ease of fabrication. To this end, high viscosity photoresist (AZ 4533) was coated to thickness of 3.5 μm over the beam as the final step in the fabrication. Optical lithography was conducted to expose the beam area with UV light. After developing the exposed photoresist, the electrostatic polymeric lensing structure with a window over the beam was created. The window size was optimized both with respect to the dimensions of the NEMS sensors and considerations of the photolithography / development steps on suspended NEMS. Wire-bonding contact pads, which are placed approximately 400 μm away from the NEMS sensor, were also opened in the same step (Figure 1). After wire bonding, the NEMS sensor including the photoresist focusing window is ready for measurement.

**Experimental Setup and the Electrospray Ionization Process.** The experimental setup shown in Supporting Figure S1 consists of the following components: (A) The ESI solution placed in a glass Hamilton syringe (Hamilton 1750 Series), (B) a syringe pump (New Era Pump Systems, Inc. NE-1000), (C) 360 μm glass tubing (LabSmith CAP360-150S), (D) a high voltage fluidic connector to supply the desired ESI voltage to the flowing ESI solution, (E) the ESI tip, (F) a copper extractor lens, (G) the NEMS device with the polymeric lensing structure.

The ESI solution was placed in a Hamilton Syringe and the ESI flow was supplied by a syringe pump at a rate of 500 nL/min. Online ESI tips with diameters of 10 μm or 30 μm (New Objective, PicoTip Emitter SilicaTip) were used. The typical ESI voltages was 5.5 kV for the high voltage fluidic connector and 1.3 kV for the extractor lens supplied by separate high voltage sources for each (High voltage fluidic connector: American Power Design, Inc. P2-600/C/Y, controlled by Agilent E3640A; Extractor: Emco High Voltage -E60, controlled by Marxlow RXN-1502D). The extractor lens (a 50x50 mm square copper plate with 15 mm diameter hole) was used at a distance (typically 4-6 mm) away from the tip. The NEMS chip was placed within 7.5 cm to 15.0 cm away from the tip. The whole setup was housed in a dry-box under atmospheric pressure. The typical

ESI parameters used in the experiments are given in Supporting Table S1. These parameters were chosen to operate in the cone-jet ESI mode and avoid any electrical discharges.

**Gold Nanoparticle Preparation.** The solutions containing 20 and 40 nm nominal diameter gold nanoparticles in citrate buffer were purchased from Sigma Aldrich (Sigma Aldrich 741965 and 741981 respectively). Using 80 mM Ammonium Acetate as the buffer (pH 7.1), serial dilutions were conducted until the final concentration was reached (Supporting Table S2). Due to the significant amount of dilution, no additional step for buffer exchange was conducted.

**Fluorescent Polystyrene Nanoparticle Preparation.** An aqueous solution containing 100 nm nominal diameter fluorescent polystyrene nanoparticles purchased from Thermo Scientific (Fluoro-Max Green Fluorescent Polymer Microspheres, CAT. NO: G100) was used in the experiments. Using 80 mM Ammonium Acetate (pH 7.1) as the buffer, serial dilutions were conducted until final concentration of $3.55\times10^9$ particles per mL was reached. Due to the significant amount of dilution, no additional step for buffer exchange was conducted.

**SARS-CoV-2 Virus Isolation.** SARS-CoV-2 infected Vero E6 cell lysate was provided by Ankara University. The samples for virus isolation were collected at the Infectious Disease Clinics, Ankara City Hospital with the official permission from Ministry of Health, Ankara City Hospital, Ethical Committee for Human Experiments (20-654, 21.05.2020). The virus strain was isolated from individuals testing positive for COVID-19 with real-time reverse transcription polymerase chain reaction (RT-PCR). Briefly, nasopharyngeal swabs were collected from positively tested individuals and promptly transferred to Dulbecco's Modified Eagle's Medium (DMEM) that was supplemented with 10% Fetal Bovine Serum (FBS) and penicillin-streptomycin (Lonza, Switzerland). The collected sample swabs were used to inoculate Vero E6 cells (ATCC, CRL-1586). Following incubation at 37º C in DMEM, the cells were monitored for cytopathic effects. Plaque assays were performed three times consecutively to purify the virus isolates.

The stock virus solution was produced by infecting Vero E6 cells in T75 cell culture flasks with the virus isolated from the three consecutive plaque assays. The degree of cytopathy was monitored and the cells were frozen once cytopathy reached 80% of the cellular monolayer. The cells were subsequently submitted to two cycles of freezing and thawing to ensure lysis. Infectivity titer showed a concentration of $2\times10^6$ PFU/mL. The cell lysate solution was then clarified by centrifugation at 1030 rcf for 15 min at 4º C (Allegra X-30, Beckman Coulter, USA). The supernatant was collected and thermally inactivated at 60º C for a period of 90 min, then preserved at -80º C for further processing.

**BoHV-1 Virus Isolation.** Bovine Herpesvirus Type 1 Cooper strain (BoHV-1 Cooper, ATCC VR864) was obtained from the departmental culture collection, Faculty of Veterinary Medicine, Department of Virology, Ankara University. The virus was cultivated in Madin-Darby Bovine Kidney (MDBK, ATCC CCL-22) cells in a similar fashion to how SARS-CoV-2 viruses were cultivated as detailed above.

Stock virus solution was obtained from MDBK cells infected with virus isolates after three consecutive plaque assays. Cells showing 80% cytopathy were subjected to 2× freeze-thaw cycles to lyse the cells, and then clarified using centrifugation. Infectivity titer showed a concentration of $1\times10^7$ PFU/mL. The virus was thermally inactivated at 60º C for a period of 90 min and preserved at -80º C for further processing.

**PEG-mediated Virus Precipitation.** Both SARS-CoV-2 and BoHV-1 underwent the same isolation and buffer exchange procedures. To selectively isolate the virus from the remainder of the cellular debris contained within the lysate, a polyethylene glycol (PEG) virus purification kit (Abcam, UK) was used. A 500 µL aliquot of the cell lysate was first diluted to 1 mL using HPLC water (Sigma-Aldrich, USA). 250 µL of the PEG-6000 solution in the precipitation kit was added to the diluted lysate and allowed to agglomerate overnight at 4º C for 16 hours. The PEG-lysate solution was then centrifuged at 3200 rcf for 30 min at 4º C (himac CT 15RE, Hitatchi, Japan). The supernatant was carefully discarded through aspiration without discarding the visible while pellet. The pellet was then resuspended in 100 µL of the resuspension solution supplied with the precipitation kit. The pellet was resuspended by trituration *via* careful pipetting.

**PEG Removal.** The PEG molecules were further removed from the purified virus mixture through salting out. 33 µL of 50 mM Tris-HCl buffer containing 4 M of KCl (pH 7.3, Bioshop, Canada) was added to 100 µL of the PEG-virus mixture. The solution was allowed to sit on ice for a period of 30 min. Once the incubation was over, the solution was centrifuged at 12,000 rcf for 10 min (himac CT 15RE, Hitatchi, Japan). The virus-containing supernatant was carefully removed without disturbing the pellet. The supernatant was further diluted in 867 µL of 80 mM Ammonium acetate buffer (pH 7.1, Merck Millipore, USA).

**Dialysis for Buffer Exchange.** In order to remove any non-volatile salts contained in the purified virus solution, dialysis was performed against ammonium acetate buffer using a Pur-A-Lyzer™ Maxi dialysis kit (12 kDa MWCO, Sigma, USA). Ammonium acetate buffer (80 mM, pH 7.1) was prepared beforehand and kept at 4º C prior to use. The PEG-purified virus solution (1 mL) was pipetted inside the dialysis tube and suspended in 1 L of ammonium acetate at 4º C for 24 hours. Following the completion of the dialysis, the sample was extracted from the dialysis tube and used as it was. The dialysis of the non-purified cell lysate was performed in a similar manner. 1 mL of the cell lysate was pipetted into a 12 kDa dialysis tube and suspended within 1 L of ammonium acetate for 24 hours at 4º C. The dialyzed cell lysate was then collected and used as it was.

**Preparation of sample solution for ESI.** For each nanoparticle and virus-containing solution prepared in ammonium acetate buffer, a volume of methanol was added up to a concentration of 10% v/v% (HPLC grade, triple filtered with 220 nm syringe filter). This was done in order to facilitate efficient electrospraying.

**Dynamic Light Scattering (DLS) measurements.** All DLS measurements were performed using the Malvern Zetasizer Nano ZS (4 mW 633 nm laser, Malvern, UK). 1 mL of the desired solution was added to a 2.5 mL quartz reusable cuvette and measured accordingly. The parameters were defined to measure proteins in water. The material refractive index was set to 1.45 for viruses, 0.2 for GNP, and 1.59 for PS, while the dispersant being water in all cases had a refractive index value of 1.33. The dispersant viscosity value used was 0.85 mPa.s for virus solutions, and 0.93 mPa.s for all other samples.[50] All measurements were run at 23º C.

**NEMS Measurements and Data Analysis.** The open-loop measurement and the PLL tracking of the first two resonance modes of the NEMS device were conducted using Lock-in-Amplifier (Zurich Instruments HF2LI) and the circuitry shown in Supporting Figure S6. MATLAB was used to extract the frequency shifts that correspond to analyte landing events using the slope of the frequency data as a criterion. Events with frequency shift values within noise level and outliers were excluded. The total NEMS device mass was calculated using dimensions acquired by SEM. The mode shape for each vibrational mode is extracted from COMSOL by fitting the resonance frequencies to experimental values. This data were then used to calculate the mass and landing position of each event and ultimately to construct mass spectra as discussed in the literature.[3] In this technique, each event is assumed a probability density function (PDF) which is then converted into PDF in the mass-position plane using the established relationships. The mean and standard deviation values for each event were calculated and used to construct histogram plots. For each event, a Gaussian function with mean and standard deviation corresponding to the specific event is assumed. These single functions are then summed up and normalized to construct the mass spectra. For diameter spectra, mass spectra were converted using the density of the particle and assuming a spherical shape.

**Nanoparticle Diameter Measurements with SEM.** In order to measure diameter of the sprayed particles, NEMS devices are imaged under SEM. Images of the particles on the chip then imported to ImageJ software. Circles are drawn over the particles and area covered by these circles are extracted. Then corresponding diameters are calculated and histogram of those calculated diameters are reported in Figure 5.

ASSOCIATED CONTENT:

**Supporting Information**.

The supporting information contains figures detailing the experimental system, nanoparticle focusing parameters, finite element simulations of NEMS devices, and droplet evaporation control runs; as well as tables for comparison of NEMS detection efficiency and the size measurements of SARS-CoV-2 with different techniques.

**Financial Interest Statement:** MSH is the founder of Sensonance Engineering Company; the other authors declare no competing interests.

AUTHOR INFORMATION

Corresponding Author

**Mehmet Selim Hanay** - Department of Mechanical Engineering, and UNAM — Institute of Materials Science and Nanotechnology, Bilkent University, 06800, Ankara, Turkey. selimhanay@bilkent.edu.tr .

**Author Contributions:**

The manuscript was written through contributions of all authors. All authors have given approval to the final version of the manuscript. ‡These authors contributed equally. M.S.H. conceived the idea; R.T.E, B.E.K., M.K., M.A, I.K. further developed the ideas for the instrumentation; H.A., B.E.K., M.A. processed the nanoparticle and virus samples; R.T.E., M.A. and B.E.K. operated the NEMS-MS system and took the data; H.S.P. , C.Y., R.T.E. and B.E.K. performed nano-fabrication; M.K., A.M.Y., B.E.K, I.K., R.T.E. and B.S. performed packaging; B.E.K., R.T.E., and H.S. P. designed the NEMS devices; Z.B.S., M.A., I.K. and M.K. performed the data analysis; B.S. and B.E.K. performed ion trajectory simulations; H.A. developed the necessary protocols for virus isolation, and performed confirmatory analysis (DLS). A.O. provided the virus samples and provided consultation on viral preparations.

**Funding Sources:** This work was supported by the Scientific and Technological Research Council of Turkey (TÜBİTAK), Grant No. EEEAG-119E503 and ERC Starting Grant REM 758769.

ACKNOWLEDGMENT

MSH acknowledges fellowship supports from TÜBA and The Science Academy, Turkey. The authors thank Sabancı University SUNUM for nanofabrication support. The authors thank Urartu Seker, Berk Kucukoglu, Uzay Tefek, Peyman Firoozy, Kadir Ulak, Yagmur Ceren Alatas, Ece Kayacilar, Enise Kartal and Sakir Duman for useful discussions.